\documentclass{elsart}

\usepackage{graphicx}
\usepackage{amssymb}

\begin{document}

\begin{frontmatter}

\title{Muon-induced neutron production \\
and detection with {GEANT4} and {FLUKA}}

\author[ICL]{H. M. Ara\'ujo\corauthref{cor1}}, 
\corauth[cor1]{Corresponding author; address: Astrophysics Group,
Blackett Laboratory, Imperial College London, SW7 2BW, UK}
\ead{H.Araujo@imperial.ac.uk}
\author[SHE]{V. A. Kudryavtsev},
\author[SHE]{N. J. C. Spooner}, 
\author[ICL]{T. J. Sumner}
\address[ICL] {Blackett Laboratory, Imperial College London, UK}
\address[SHE] {Department of Physics \& Astronomy, University of Sheffield, UK}

\begin{abstract}
We report on a comparison study of the Monte Carlo packages GEANT4 and
FLUKA for simulating neutron production by muons penetrating deep
underground. GEANT4 is found to generate fewer neutrons at muon
energies above $\sim$100~GeV, by at most a factor of 2 in some
materials, which we attribute mainly to lower neutron production in
hadronic cascades. As a practical case study, the muon-induced neutron
background expected in a 250~kg liquid-xenon WIMP dark matter detector
was calculated and good agreement was found for the recoil event
rates. The detailed model of neutron elastic scattering in GEANT4 was
also shown to influence the nuclear recoil spectrum observed in the
target, which is presently a shortcoming of FLUKA. We conclude that
both packages are suited for this type of simulation, although further
improvements are desirable in both cases.
\end{abstract}


\end{frontmatter}

% main text %%%%%%%%%%%%%%%%%%%%%%%%%%%%%%%%%%%%%%%%%%%%%%%%%%%%%%%%%%%

%%%%%%%%%%%%%%%%%%%%%%%%%%%%%%%%%%%%%%%%%%%%%%%%%%%%%%%%%%%%%%%%%%%%%%%
\section{Introduction}

The knowledge of neutron fluxes in underground laboratories is crucial
in experiments searching for and detecting rare events associated with
neutrino interactions, double-beta decay and dark matter WIMPs (Weakly
Interacting Massive Particles). For example, in dark matter searches,
in order to reach 10$^{-10}$~pb sensitivity to the WIMP-nucleon
cross-section, the neutron flux produced by muons underground must be
known and suitably attenuated. Other sources of neutrons can, in
principle, be mitigated by using radio-pure materials and components
together with appropriate shielding. In many existing underground
laboratories (Gran Sasso, WIPP, Kamioka, Modane, Boulby and others)
the muon-induced neutron flux may be a key limiting factor for
detector sensitivity and may require an active veto system to be
installed around the target. This can easily be the dominant cost in a
large detector system.

Atmospheric muons, generated in Extensive Air Showers at high
altitudes in the atmosphere, penetrate deep underground and produce
neutrons in the rock surrounding a laboratory or in the materials
inside it. Most neutrons do not come directly from muon-nucleus
spallation reactions, but from the muon-initiated hadronic and
electromagnetic cascades. These neutrons spread away from the primary
muon track, thermalising slowly, and finally get captured by protons
or other nuclei. Some, however, can strike a nucleus in the target
volume transferring sufficient energy to be detected. This is an
irreducible background for many underground experiments. In dark
matter searches, for example, the single nuclear-recoil signals from
neutron elastic scattering cannot be distinguished from WIMP
interactions. An increased gamma activity of some materials is also
expected from neutron activation.

Underground neutron fluxes are difficult to measure since they are
overwhelmed by gamma-rays from local radioactivity. In addition, most
neutrons at the aforementioned underground laboratories come from rock
radioactivity: muon-induced neutrons contribute typically less than
one percent to the total, unshielded neutron flux. Although these
neutron events may have a clear signature of an associated muon and/or
cascade, large detectors are needed to study them with sufficient
accuracy, especially to understand the neutron behaviour at large
distances from the muon track. Monte Carlo (MC) simulations are thus
crucial in designing high-sensitivity detectors. Naturally, this
reliance on simulation alone must be accompanied by validation of the
codes against other experimental data, and by comparison between
different packages.

FLUKA \cite{fluka} is a well-established simulation tool in nuclear
and particle physics, and has been used for neutron calculations in a
variety of experiments. However, it does not treat the elastic
scattering of neutrons at low energies with sufficient detail (on an
event-by-event basis) for the purpose of dark matter
experiments. So-called `Kerma' factors are used to generate energy
deposition from nuclear recoils (other than protons), which are
equivalent to the average recoil energy for a certain neutron
energy. The GEANT4 \cite{geant4} toolkit can potentially be used for
end-to-end simulations of experiments, from background calculations
down to detailed detector characterisation. Its object-oriented design
and open-source nature make it rather flexible. Critically, it
generates and tracks in a realistic way the recoiling nuclei from
individual neutron elastic interactions.

In this paper we compare GEANT4 and FLUKA simulations of neutron
production, and validate them against available experimental data.
Some FLUKA results have already been published in
Ref.~\cite{kudryavtsev03} (hereafter `Paper~1') and Ref.~\cite{wang01}
(Paper~2).  We believe this to be a useful test of GEANT4 physics for
low-background experiments, an area where it finds increasing
application. We describe the full simulation of the neutron background
in an idealised WIMP dark matter detector, hypothetically located at
the Boulby mine (2.8~km~w.e.) --- location of the UK Dark Matter
Collaboration experiments. In particular, we compare the underground
neutron spectrum, veto and shielding efficiencies and nuclear recoil
spectrum expected in a 250 kg xenon target with the FLUKA results
published in Ref.~\cite{carson04} (Paper~3). Muon propagation from the
surface is not addressed here. Instead, the MC code MUSUN
\cite{kudryavtsev03} was used to calculate the muon energy spectrum
and angular distribution at Boulby in the way described in Paper~3.

The present simulations made use of GEANT4 release 6.2 and
FLUKA-2003. The FLUKA modelling described in Paper~3 was performed
with FLUKA-2002, whereas the results of Paper~1 and Paper~2 were
obtained using FLUKA-1999.

%%%%%%%%%%%%%%%%%%%%%%%%%%%%%%%%%%%%%%%%%%%%%%%%%%%%%%%%%%%%%%%%%%%%%%%
\section{Muon-induced neutron production}

Since GEANT4 offers alternative models to treat certain physics
processes, we mention here briefly which ones were used in these
simulations. The physics implemented in the toolkit is fully described
in Ref.~\cite{physrefman}. Muon-induced spallation (or muon
photonuclear interaction) is modelled above 1~GeV muon energy; the
final-state generator relies on parameterised hadronic models. Gamma
inelastic scattering, the real photonuclear interaction, generates its
hadronic final states using a chiral-invariant phase-space decay model
below 3~GeV; a theoretical quark-gluon string model simulates the
punch-through reaction at higher energies. The hadronic interaction of
nucleons, pions and kaons is simulated with the quark-gluon string
model above 6~GeV, an intra-nuclear binary cascade model at lower
energies and a pre-equilibrium de-excitation stage below 70~MeV. The
equilibrium stage considers fragment and gamma evaporation, fission,
Fermi break-up and multi-fragmentation of highly-excited nuclei.
Neutron transport and interactions are described by data-driven models
below 19~MeV. 

The choice of electromagnetic physics is more straightforward and will
not be described here, except to say that the so-called `Low Energy'
package was used for $e^-$ and gammas interactions, and the standard
models were applied to the remaining particles (including muons). The
production thresholds (`cuts') considered in these simulations were a
few tens of keV for gammas and $\sim$1~MeV for $e^-$ and $e^+$ in all
materials. No thresholds were applied to neutron tracking.

\subsection{Neutron production in hydrocarbon scintillator}

In order to study neutron production by muons in several materials we
consider a $\mu^-$ beam incident at the centre of a block with
thickness 3200~g/cm$^2$ and comparable transverse dimensions. We
compared the muon energy spectra at the end of the block from GEANT4,
FLUKA and the muon propagation code MUSIC \cite{music} and found them
to be in good agreement \cite{idm04}.  This validates the methods used
to calculate muon interaction cross-sections.

Muon-induced cascades require a certain length of material to develop
and reach equilibrium in the number of neutrons produced per unit muon
track length. To avoid this and other edge effects, only neutrons
produced in the central half-length of the block are considered. In
counting the number of neutrons created in a cascade one must avoid
double-counting inelastically scattered ones, which often share the
final state with newly generated neutrons. In the GEANT4 simulation we
consider that the highest energy one in the final state corresponds to
the initial neutron\footnote{Strictly, the final state of the
inelastic interaction has no memory as to which neutron corresponds to
the incident one, and so this assertion does not always hold. We note,
however, that this will not affect the neutron multiplicity (and hence
the total yield), only the production spectrum.} and remove it from
the count (akin to the concept of `stars' in FLUKA). For muon energies
below $\sim$100~GeV one must also correct for muon energy loss, since
the neutron yield decreases noticeably as the muon progresses through
the material. These cases are corrected by extrapolating the trend of
the decreasing yield to the beginning of the muon beam.

\begin{figure}
   \includegraphics[width=8.5cm]{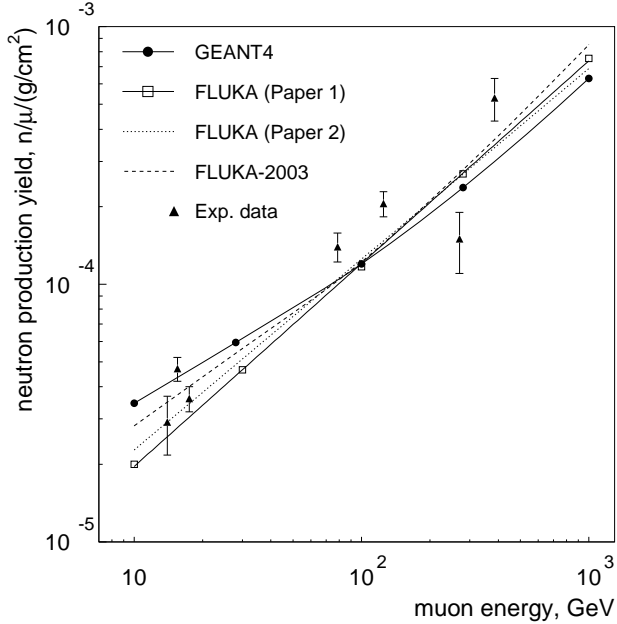}
   \caption{Dependence of the neutron yield per unit muon track length
   on muon energy for C$_{10}$H$_{20}$ scintillator. The experimental
   data represent measurements at depths varying between 20~m~w.e. and
   5200~m~w.e. with the corresponding underground muon spectrum. Here
   they are plotted as a function of the mean muon energy. This
   approximation is justified in Paper~1 and Paper~2, where
   references to the original experiments can be found. Note the
   higher neutron production rate at low energies in present
   (FLUKA-2003) simulations compared to Paper~1 and Paper~2.}
   \label{scintillator}
\end{figure}

A generic hydrocarbon with composition C$_n$H$_{2n}$ (density
$\rho$=0.8~g/cm$^3$) was the lightest material simulated in this
study. This can represent commonly used neutron-shielding hydrocarbons
as well as typical organic scintillators (e.g. C$_{10}$H$_{22}$) for
which experimental data are available. The total neutron yields, which
include neutrons of any energy, are shown in Fig.~\ref{scintillator}
as a function of muon energy (the MC statistical uncertainties are
comparable to the size of the data markers). Although GEANT4 results
agree, at higher energies, with the power law $E^{0.74-0.79}$
predicted by FLUKA in Papers~1 and 2, there is an enhancement with
decreasing energy relative to the previous FLUKA simulations.  Present
results with FLUKA-2003 show similar enhancement of neutron production
at low energies over the simple power law. Both GEANT4 and FLUKA are
consistent with the experimental data shown, which have been measured
at various depths around the world, keeping in mind the spread in
experimental results.

\begin{figure}
   \includegraphics[width=8.5cm]{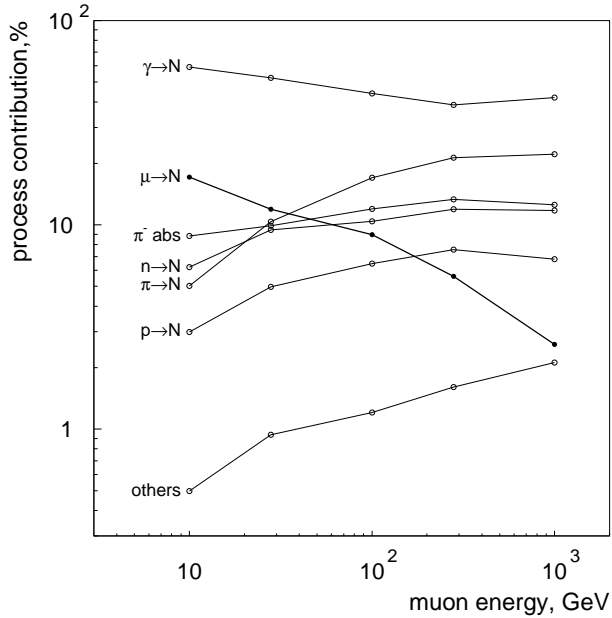}
   \caption{Relative contribution of individual processes to the total
   neutron yield in scintillator from the GEANT4 simulation. The
   processes shown explicitly are photonuclear interaction of gammas
   ($\gamma\!\rightarrow\!N$), muon spallation
   ($\mu\!\rightarrow\!N$), proton spallation ($p\!\rightarrow\!N$),
   pion ($\pi^+$ and $\pi^-$) spallation ($\pi\!\rightarrow\!N$),
   $\pi^-$ absorption at rest ($\pi^-$~abs) and neutron inelastic
   scattering ($n\!\rightarrow\!N$). The processes gathered under
   `others' include electronuclear reactions
   ($e^\pm\!\rightarrow\!N$), kaon spallation and K$^-$ absorption at
   rest, as well as spallation reactions involving light fragments
   ($^2$H, $^3$H, $^3$He and $\alpha$-particles), anti-nucleons
   ($\bar{n}$, $\bar{p}$) and short-lived hadrons
   ($\Lambda$,$\Sigma$,$\Xi$, etc.).}
   \label{processes1}
\end{figure}

We have also analysed the importance of individual neutron-producing
processes in this simulation, which can be compared to similar data
for FLUKA presented in Papers~1 and 2. The relative contributions of
the most important mechanisms to the total neutron rate are shown in
Fig.~\ref{processes1} for GEANT4. There is a degree of agreement
between the two codes, in that both predict neutron production in
electromagnetic cascades (real photonuclear interaction) to dominate
at lower energies and to decrease in importance with increasing muon
energy, and generation in hadronic cascades to become more important
with increasing muon energy. Both codes confirm that most neutrons are
not produced in direct muon-induced spallation, but rather in the
cascades muons initiate, and more so at higher energies. However, the
GEANT4 results reveal a greater dominance of electromagnetic cascades
at low energies (higher total neutron yield,
cf. Fig.~\ref{scintillator}), and this scenario is not significantly
different at high energies, where FLUKA predicts neutron production in
hadronic cascades to take over. In this material, although GEANT4
appears to overproduce neutrons in electromagnetic cascades, it
underproduces in hadronic cascades compared to FLUKA, giving similar
total yield in both codes. It should be noted that this could be due
to a difference in the total interaction cross-sections, in the
neutron-production cross-sections or in the final-state multiplicity
of secondary particles.

\begin{figure}
   \includegraphics[width=8.5cm]{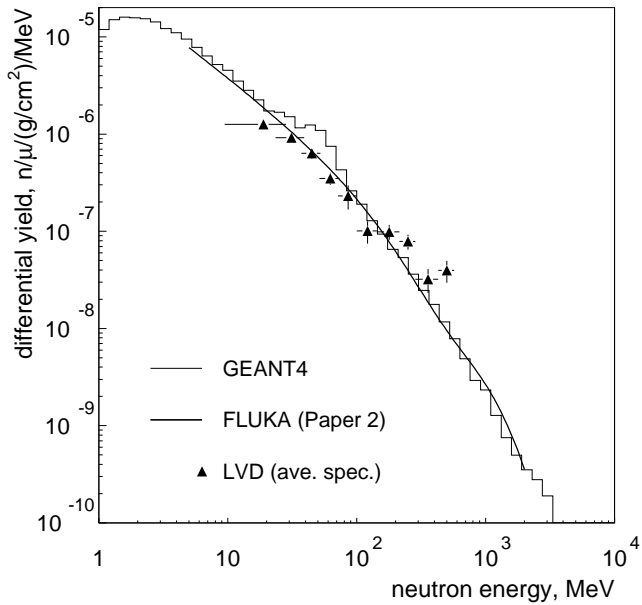}
   \caption{Differential energy spectrum of muon-induced neutrons
   produced in scintillator. The GEANT4 absolute yield is compared
   with the relative FLUKA spectrum represented by the
   parameterisation given in Paper~2. Experimental data refer to an
   average neutron spectrum in the scintillator volume as measured by
   the LVD experiment \cite{aglietta99}. The FLUKA and LVD datasets
   were normalised to the GEANT4 spectrum for visual agreement.}
   \label{energy}
\end{figure}

The neutron energy spectrum at production in the scintillator,
calculated in the way described above for the total neutron yield, is
shown in Fig.~\ref{energy} for 280~GeV muons, a value close to the
average muon energy at several underground labs. These data are
compared with a parameterisation of the relative spectrum given in
Paper~2 for FLUKA. Measurements with the LVD detector
\cite{aglietta99}, located at the Gran Sasso underground laboratory
(mean depth of 3.6 km w.e.), are also shown --- normalised for visual
agreement. Although these experimental data do not show the neutron
energy spectrum at production, but a volume-averaged spectrum instead,
they are nevertheless a useful guide. Note that the parameterisation
of the neutron spectrum given in Paper~2 does not show a feature above
$\sim$10 MeV due to the Giant Dipole Resonance (GDR) and the so-called
`Quasi-Deuteron' region. This feature is, however, also seen in the
FLUKA spectrum, averaged over the entire scintillator volume,
presented in Paper~1.

\begin{figure}
   \includegraphics[width=8.5cm]{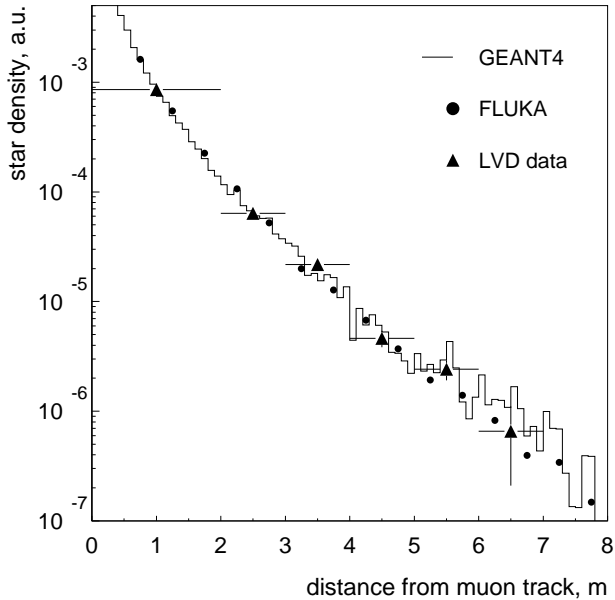}
   \caption{Lateral distribution of neutron inelastic interactions
   from the primary muon track for the muon spectrum at about
   2.8~km~w.~e. (280 GeV mean energy). The FLUKA `star density' is
   compared with LVD data \cite{aglietta99} and with a similar
   quantity calculated with GEANT4. Normalisation is done to the
   GEANT4 distribution for visual agreement at small distances.}
   \label{lateral}
\end{figure}

Neutrons can be created some distance away from the primary muon
track, an issue to consider when modelling veto performances. The
FLUKA `star density' output is an estimator for the spatial
distribution of neutron reactions, which can be compared to
experimental measurements carried out with large scintillator volumes
--- as reported in Paper~1. A FLUKA `star' is a hadronic inelastic
interaction (neutron spallation, in this instance) above a certain
energy threshold. An equivalent quantity was calculated from the
GEANT4 dataset, which was obtained for the input muon spectrum
expected at 2.8~km~w.e. underground. The result is shown in
Fig.~\ref{lateral}.  Both simulations agree well with the LVD
measurement \cite{aglietta99}. More recent LVD data \cite{menghetti04}
suggest a higher number of neutrons at large distances from the muon
track.  Note, however, that LVD is not a detector with a single
uniform medium but has a modular structure with gaps between
modules. Thus any such comparison should be considered with caution.

\subsection{Neutron yields in other materials}

\begin{figure}
   \includegraphics[width=8.5cm]{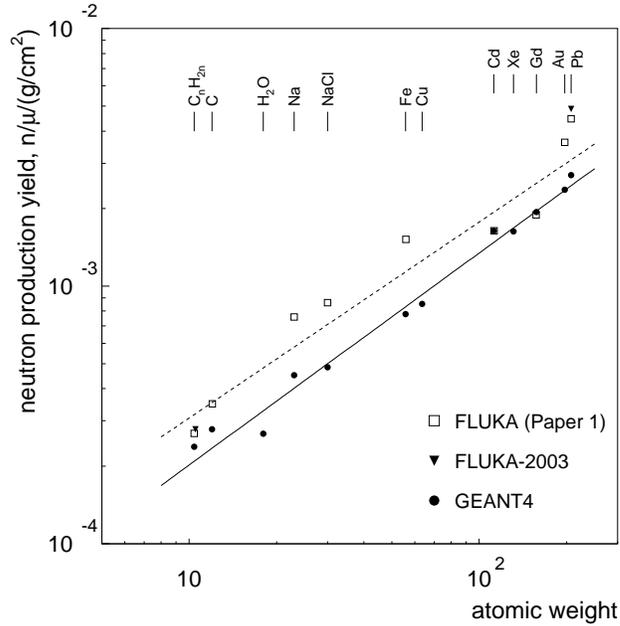}
   \caption{Dependence of the neutron yield on the average atomic
   weight of the material in FLUKA (Paper~1) and GEANT4 (this work)
   for 280~GeV muons, for the elements and compounds indicated. The
   power-law parameterisations are given in the text. Two data-points
   show the present FLUKA-2003 results for scintillator and lead.}
   \label{weight}
\end{figure}

Neutron production in several other materials of interest to low
background experiments was studied with GEANT4 for a fixed muon energy
of 280~GeV, using the block geometry and the neutron event selection
rules described previously for organic scintillator. The variation of
the total neutron yield with atomic weight of the element (or average
atomic weight for compounds) is shown in Fig.~\ref{weight}, along with
the FLUKA results obtained in Paper~1 (we also show new results for
scintillator and lead from FLUKA-2003 simulations). The two codes
differ at most by a factor of 2 --- with FLUKA predicting consistently
higher rates at this energy --- but the largest discrepancies occur
precisely for materials of great interest for underground experiments:
notably NaCl (the rock at WIPP and Boulby), Fe and Pb (shielding
materials). The statistical uncertainty associated with each data point
is much smaller than the discrepancy between corresponding FLUKA and
GEANT4 data points. The rates can be fitted by a power-law dependence
of the atomic weight $A$: $R=bA^\beta$. We obtain
$b$=(3.0$\pm$0.4)$\times 10^{-5}$ and $\beta$=0.82$\pm$0.03 for the
GEANT4 simulation, while Paper~1 reports $b$=(5.33$\pm$0.17)$\times
10^{-5}$ and $\beta$=0.76$\pm$0.01 for FLUKA.

\begin{figure}
   \includegraphics[width=8.5cm]{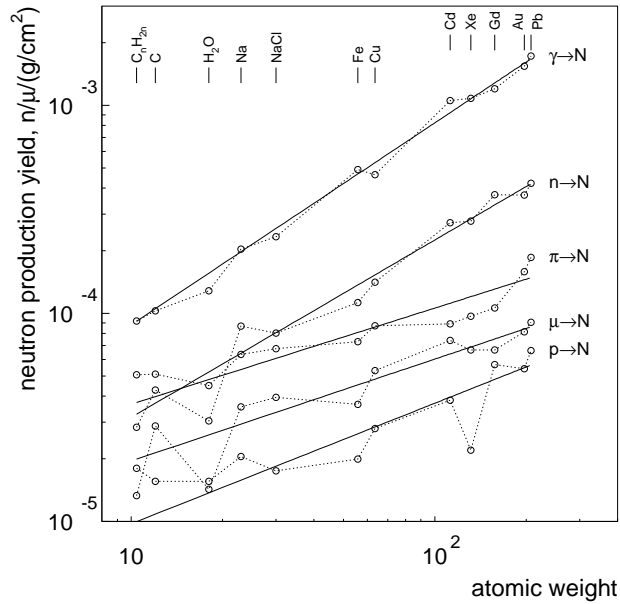}
   \caption{Contribution of different processes to the neutron yield
   as a function of atomic weight of the material for the GEANT4
   simulation of 280~GeV $\mu^-$. The processes include
   photoproduction ($\gamma\!\rightarrow\!N$), neutron inelastic
   scattering ($n\rightarrow\!N$), $\pi^+$ and $\pi^-$ spallation
   ($\pi\!\rightarrow\!N$), muon spallation ($\mu\!\rightarrow\!N$)
   and proton spallation ($p\!\rightarrow\!N$).}
   \label{processes2}
\end{figure}

Valuable information can be gained by decomposing the total neutron
rates into the originating physical processes and studying how these
vary with atomic weight. The data in Fig.~\ref{processes2} break down
the contribution of five inelastic scattering processes at 280~GeV
muon energy. Real photonuclear interactions clearly dominate in all
materials, increasing linearly with $A$. This is indeed the dependence
expected for the total cross-section, suggesting that the number of
neutrons produced per reaction remains approximately constant ---
confirmed by the GEANT4 team \cite{hp}. Photoproduction in FLUKA gives
less neutrons in scintillator ($6.5 \times 10^{-5}$ compared to $9.2
\times 10^{-5}$ n/$\mu$/(g/cm$^2$) in GEANT4) and more neutrons in
lead ($2.4 \times 10^{-3}$ and $1.7 \times 10^{-3}$
n/$\mu$/(g/cm$^2$), respectively). Secondary neutron production in
GEANT4, resulting from neutron inelastic scattering, increases
slightly more slowly ($A^{0.85}$) than photoproduction. Yields from
pion ($\pi^+$, $\pi^-$), proton and muon spallations vary
approximately as $A^{0.5}$. The scattered nature of the data for a
given process is not statistical in origin. The relative error for
direct muon spallation (which also governs hadronic cascade
statistics), is only a few percent.

In conclusion, muon-induced neutron production in GEANT4 is generally
lower than that in FLUKA at high muon energies, although this
difference is typically smaller than a factor of 2 --- an acceptable
agreement for many purposes. This is due to a generalised neutron
deficit from hadronic cascades. This underproduction with respect to
FLUKA may be expected, since the latter does not model the emission of
fast nuclear fragments from highly-excited nuclei
(multi-fragmentation), especially important in heavy nuclei, thus
leaving more energy available for neutron evaporation \cite{hp}. In
heavy elements electromagnetic cascades also underproduce neutrons
with respect to FLUKA but this is not the case for the light
hydrocarbon scintillator. Direct production in muon-nucleus
interactions is never dominant. Nevertheless, we analyse this process
more closely in the following section.

Experimental data for neutron production by energetic muons in heavy
targets is scarce and not straightforward to interpret without detailed
modelling, given the complicated experimental set-ups and poor
knowledge of the neutron detection efficiencies. In
Ref.~\cite{wulandari04} FLUKA simulations of neutron production by
high-energy muons were compared to experimental measurements in lead
\cite{gorshkov74,bergamasco73}. Some of these results relevant to high
muon energies, converted to the neutron yield per muon are:
$1.75\times 10^{-3}$~n/$\mu$/(g/cm$^2$) for a mean muon energy of
$\simeq$110~GeV (depth 800~m~w.e.) \cite{gorshkov74} and $1.16\times
10^{-2}$~n/$\mu$/(g/cm$^2$) for a mean muon energy of $\simeq$310~GeV
(depth 4300~m~w.e.). We used the muon energy dependence of the neutron
yield ($\propto E^{0.79}$ in hydrocarbon \cite{kudryavtsev03}) to
convert the quoted neutron yields to those at 280~GeV and to compare
them to our simulations. We obtained $\sim\!3.7 \times
10^{-3}$~n/$\mu$/(g/cm$^2$) from Ref.~\cite{gorshkov74} and $\sim\!1.1
\times 10^{-2}$~n/$\mu$/(g/cm$^2$) from Ref.~\cite{bergamasco73}. The
first result is in good agreement with the MC predictions (especially
FLUKA's) --- assuming that the total neutron yield was measured in the
experiments --- whereas the second clearly indicates a higher yield
than the simulations. Moreover, the quoted target thickness for these
experiments is too small for the muon-induced cascades to develop
fully, thus reducing the number of detected neutrons and making direct
comparison difficult. In conclusion, some experiments may suggest that
both FLUKA and GEANT4 underestimate neutron production in heavy
targets, although more conclusive measurements are required.

\subsection{Neutron production by muon-induced spallation}

The inelastic scattering of muons off nuclei can be explained in terms
of the exchange of quasi-real virtual photons --- hence it is also
known as the muon photonuclear interaction. Presently, the GEANT4
treatment of the muon-nucleus interaction uses the total cross-section
formulae given in Ref.~\cite{borog} and Ref.~\cite{bezrukov} and the
parameterisation of the real photonuclear cross-section
($\gamma\!-\!N$) from Ref.~\cite{caldwell}. This treatment applies to
muon energies above 1~GeV and a minimum energy transfer of 200~MeV
\cite{physrefman}. The final-state hadronic vertex uses parameterised
hadronic models. Although FLUKA follows an approach similar to that
currently adopted in GEANT4, it is nevertheless useful to compare the
two models, and these with experimental data.

\begin{figure}
   \includegraphics[width=8.5cm]{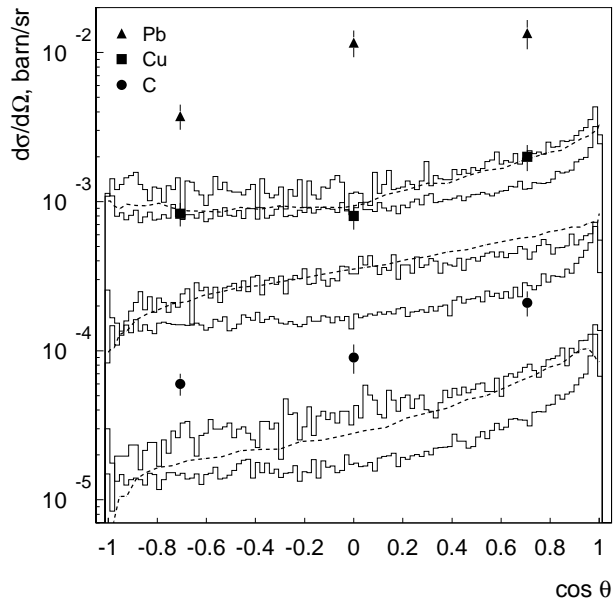}
   \caption{Differential cross-section of neutron production by
   190~GeV muons for a 10~MeV threshold in neutron energy. The data
   points represent the results of the NA55 experiment. The thin-line
   histogram shows the GEANT4 simulation considering muon-nucleus
   interaction only; the thick histogram includes all physics
   processes. The dashed line represents the FLUKA results for the
   latter case.}
   \label{spallation}
\end{figure}

The CERN NA55 has produced relevant data for such a comparison. In
this experiment analysis was performed of the production of fast
neutrons by 190~GeV muons in graphite, copper and lead
\cite{chazal02}. Given the aim in NA55 of studying neutron production
in direct muon-nucleus interactions, small targets were used in order
to minimise secondary reactions. Three neutron detectors were located
at 45$\mathrm{^o}$, 90$\mathrm{^o}$ and 135$\mathrm{^o}$ from the muon
beam, for which an energy threshold (in neutron energy) of
$\simeq$10~MeV was indicated. Simulations were thus set-up with FLUKA
and GEANT4 to model the three targets. The angular distribution of
neutrons emitted from a target was recorded in order to calculate the
differential cross-section for neutron production, which is shown in
Fig.~\ref{spallation}. GEANT4 results are given for two cases: firstly
-- data including only muon-induced spallation, and secondly -- data
from allowing the complete set of physics processes.  The FLUKA
simulation modelled only the latter case.

Several conclusions can be drawn from the data shown. Firstly, if
either MC is to be believed, then the experiment could not measure
direct muon-induced spallation exclusively. The contribution of
secondary reactions cannot be neglected, especially for the carbon and
copper targets. Secondly, both codes agree quite reasonably, in fact
better than anticipated given the results presented in the previous
section. Therefore the discrepancy observed for Pb in
Fig.~\ref{weight} does not arise from muon spallation --- in fact,
GEANT4 predicts a slightly higher fast-neutron yield than FLUKA for
this element, as shown in Fig.~\ref{spallation} --- but is instead
associated with cascades (the main contribution in Fig.~\ref{weight}),
which do not build up in thin targets. Finally, it is apparent that
the experimental points lie far above the MC data. If the disagreement
for the light materials is already unreasonable, the case of lead
cannot be easily explained. In fact, even if a much lower neutron
threshold of 1~MeV is considered, both MC models would still predict a
lower cross-section than those reported in Ref.~\cite{chazal02}.

The production of lower energy neutrons by deep inelastic scattering
(DIS) of 470~GeV muons in lead was studied by the E665 Collaboration,
who found average neutron multiplicities per DIS event of $\simeq$5
for neutron energies under 10~MeV \cite{adams95}. A value of 3.7 is
obtained for the GEANT4 simulation of the $\mu$--N process at this
muon energy, in reasonable agreement with the experimental result. The
simulated neutron spectrum (per unit energy) exhibits a
double-exponential behaviour below 10~MeV --- also in agreement with
the experimental findings. The two decay constants, due to neutron
evaporation from the thermalised nucleus and from pre-equilibrium
emission, are characterised by nuclear temperatures of 0.93~MeV and
3.7~MeV, compared to 0.7$\pm$0.05~MeV and 5$\pm$1~MeV obtained in
E665. In conclusion, the spallation of neutrons under 10~MeV as
predicted by GEANT4 for lead does not conflict with these experimental
data.

The role of the minimum energy transfer in the muon photonuclear
models in neutron production was pointed out in Paper~2. This
threshold comes about because the virtuality of the photon can no
longer be neglected when it becomes comparable to its
energy. Recently, the total $\mu\!-\!N$ cross-section was reported to
increase by 2--3 times if the minimum energy transfer is decreased
from 140~MeV to 10~MeV, based on the parameterisation used in FLUKA
\cite{gerbier04}. We have confirmed that this difference is only
10-15\% greater for the 200~MeV threshold in GEANT4. The
aforementioned study also found that the parameterisation used to
describe the $\gamma\!-\!N$ cross-section in FLUKA \cite{bezrukov}
(similar to that from Ref.~\cite{caldwell} used in GEANT4)
overestimates more rigorous theoretical calculations when extrapolated
to low energy gammas. Consequently, the increase in the muon
cross-section with decreasing threshold is not expected to be as large
as mentioned above. In any case, as pointed out in Paper~2, we expect
many more neutrons to be produced by bremsstrahlung (real) photons
with low energies in electromagnetic cascades than by virtual ones in
muon interactions with small energy transfers.

%%%%%%%%%%%%%%%%%%%%%%%%%%%%%%%%%%%%%%%%%%%%%%%%%%%%%%%%%%%%%%%%%%%%%%%
\section{Underground neutron fluxes: a case study}

The UK Dark Matter Collaboration (UKDMC) has been assessing the
feasibility of a xenon-based tonne-scale dark matter experiment to be
installed at the Boulby Underground Laboratory. In this context,
initial calculations using FLUKA, reported in Paper 3, have so far
been performed of the muon-induced background in a 250~kg xenon
target. Building on that work we present here a case-study comparison
between FLUKA and GEANT4. The calculated neutron fluxes and spectra at
the rock/cavern boundary and after various shields are also relevant
to other underground experiments in different laboratories.

\subsection{Energy spectrum at rock surface}

In these simulations an air-filled cavern (otherwise empty) with a
height of 5~m and horizontal dimensions of $6\times 6$~m$^2$ was
located in a $20\times 20\times 20$~m$^3$ cube of pure NaCl
($\rho$=2.2~g/cm$^3$), at a depth of 10~m from the top. NaCl
represented the halite rock surrounding the laboratory. Muons were
sampled on 5 sides of the salt cube, according to the energy spectrum
and angular distribution calculated with the MUSUN Monte Carlo code
for a rock overburden of 2800~m~w.e. \cite{kudryavtsev03}. The muon
flux experimentally measured underground, used to normalise the
simulations, was $(4.09\pm 0.15)\times 10^{-8}$~muons/cm$^2$/s
\cite{kudryavtsev02,robinson03}.

\begin{figure}
   \includegraphics[width=8.5cm]{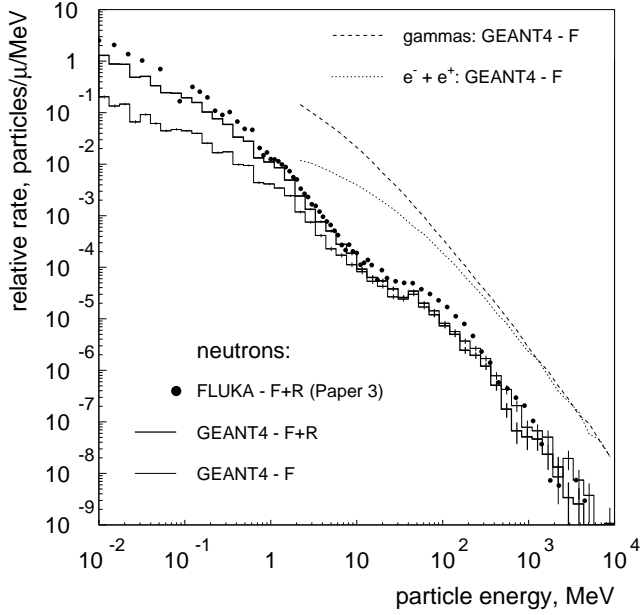}
   \caption{Neutron energy spectrum at the rock/cavern boundary
   obtained with FLUKA and GEANT4. The curve labelled `F' includes
   only those neutrons entering the cavern for the first time; the
   curve marked `F+R' accounts also for neutrons reflected back from
   the walls after crossing the empty laboratory. Also shown are the
   differential spectra for gammas and electrons ($e^-$ and $e^+$) at
   the rock face.}
   \label{rockface}
\end{figure}

The GEANT4 and FLUKA results for the relative neutron rate (the number
of neutrons per muon entering the cavern) are plotted in
Fig.~\ref{rockface} for two cases: i) for neutrons crossing the
interface for the first time and ii) also including those re-entering
the laboratory after being scattered back from the rock. For the
latter case, results agree within a factor of 2 for most of the energy
range plotted, which is consistent with the level of agreement of the
total yields found previously for 280~GeV muons. In the range
1--10~MeV --- most interesting for a WIMP dark matter experiment ---
the agreement is even better.  The integral neutron flux above 1~MeV
is $6.9\times 10^{-10}$~n/cm$^2$/s according to GEANT4, close to the
FLUKA prediction of $8.7\times 10^{-10}$~n/cm$^2$/s.

The GEANT4 simulation also provided the gamma and electron/positron
fluxes at the rock face, which are shown in Fig.~\ref{rockface}.
These data include all muon events, not just those producing neutrons
in the cavern. Many more gammas than neutrons enter the experimental
hall. In fact, most neutron events are accompanied by gammas (and
$e^\pm$) --- assuming a minimum energy of 1~MeV for both neutrons and
gammas. Together with the primary muons, this electromagnetic
component can help to tag neutrons emerging from the rock.

If a veto detector were to cover the entire surface of the laboratory
walls, approximately 70\% first-time (F) neutrons above 1~MeV would
register a coincident muon signal. All neutrons arrive within
$\simeq$0.1~$\mu$s of the primary muon if no reflections are
considered. Of the remaining (muon-unvetoed) neutron events, a further
one-half contain significant ($>$1~MeV) electromagnetic energy in the
form of gamma rays crossing the boundary which can help veto the
neutron. In total, only $\simeq$12\% of neutron events would lack a
coincident veto signal, and may therefore be difficult (or impossible)
to detect. When reflected neutrons are included in this analysis
(F+R) the latter fraction decreases to 8\%, as some neutrons
thermalise enough to undergo radiative capture, making their detection
easier. In this, more realistic case, neutrons above 1 MeV enter the
cavern up to $\simeq$1~$\mu$s after the primary muon, whereas lower
energy neutrons can take as long as 1~ms.

\subsection{Shielding efficiency}

\begin{figure}
   \includegraphics[width=8.0cm]{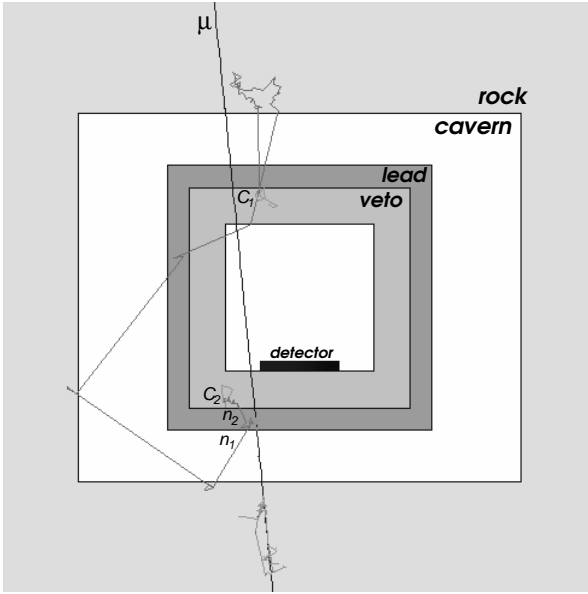}
   \caption{Monte Carlo geometry model for a 250~kg liquid xenon
   module shielded by 50~cm of hydrocarbon scintillator (40~g/cm$^2$)
   and 30~cm of lead (340~g/cm$^2$). The figure also shows a GEANT4
   event in which a 280~GeV muon produces two neutrons ($n_1$, $n_2$)
   in the lead shielding which are captured in the hydrocarbon veto
   ($C_1$, $C_2$).}
   \label{detector}
\end{figure}

\begin{figure}
   \includegraphics[width=8.5cm]{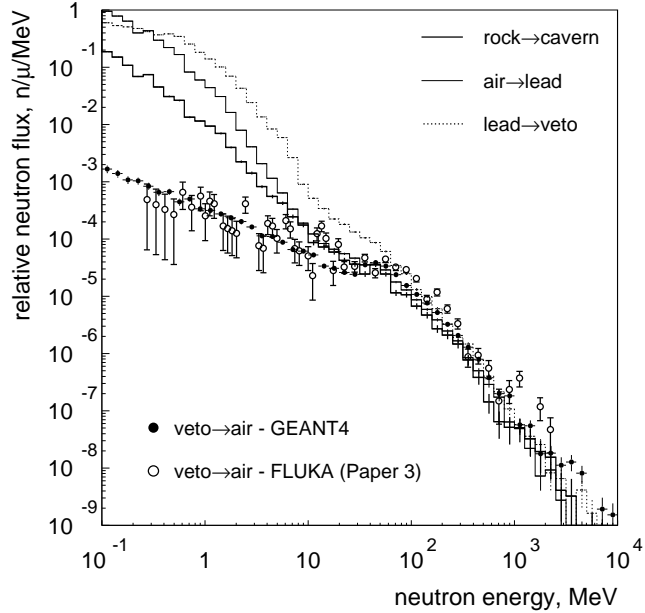}
   \caption{Energy spectra of muon-induced neutron fluxes across
   several boundaries.
   The thick line represents the flux at the rock face for an empty
   cavern (curve labelled `GEANT4 - F+R' in Fig.~\ref{rockface}). The
   spectrum of neutrons exiting the hydrocarbon veto obtained with
   FLUKA (Paper~3) is also shown.}
   \label{shielding}
\end{figure}

In the next simulation a shielded detector module is added to the
empty cavern, as shown schematically in Fig.~\ref{detector}. A 250~kg
liquid xenon target, contained in a copper vessel 2~cm thick, is
surrounded by 50~cm of C$_{10}$H$_{20}$ scintillator to form an active
veto (inner dimensions 2~m$\times$2~m$\times$2~m) followed by 30~cm of
lead shielding. The neutron spectra across various boundaries in this
set-up are shown in Fig.~\ref{shielding}. The figure confirms that
most neutrons are produced in the lead shielding --- either directly
created by muons, as shown in Fig.~\ref{detector}, or by high-energy
gammas and hadrons coming from the cavern walls or generated in the
shielding itself. Although the hydrocarbon is very efficient in
attenuating the neutron flux (by moderation and absorption), this
lead/hydrocarbon combination provides effective shielding only for
neutrons below 10~MeV. The FLUKA results presented in Paper~3 show
similar trends. It should be noted that in that work the neutron
spectrum at a particular boundary was calculated before a new layer is
added to the geometry, e.g. the effect of the lead shielding is
calculated before an inner hydrocarbon layer is introduced (and so the
difference to the one shown here is greater than a factor of $\simeq$2 for
the total production rate in Fig.~\ref{weight}). Similarly, the
neutron flux exiting the hydrocarbon-based veto was calculated before
the detector is added to the model. Given the small size of the
detector itself compared to the inner surface of the veto, the
comparison shown in Fig.~\ref{shielding} is appropriate. The neutron
spectra emerging across the veto calculated by the two MCs are in good
agreement, especially for neutron energies below a few tens of MeV.

\subsection{Neutron detection in veto and xenon target}

The prompt scintillation signal is the detection mechanism now
considered for the liquid xenon target. To account for quenching of
the scintillation yield for nuclear recoils relative to
electron-recoil interactions, a quenching factor of 0.2 is applied to
the energy deposited in neutron elastic scattering events
\cite{akimov02,arneodo00}. We assign a detection threshold of 2~keV
electron-equivalent (e.e.) or `visible' energy to the target, which is
a realistic scenario for xenon WIMP detectors. In the preceding
analysis the hydrocarbon scintillator had the role of passive
shielding. We now consider it as an active veto, which is a much more
powerful way to reduce the neutron background in the xenon target. A
threshold of 100~keV e.e. is assumed for detection of proton recoils
(and other particles)
in a typical veto using this type of organic scintillator (500--700
keV proton recoil energy \cite{anghinolfi79}). Such an idealised veto
system, with full angular coverage, can reach high rejection
efficiency \cite{smith04,carson04b} of neutrons produced in detector
components, by detecting proton recoils and gamma rays from radiative
neutron capture --- two such capture events are illustrated in
Fig.~\ref{detector}.

\begin{figure}
   \includegraphics[width=8.5cm]{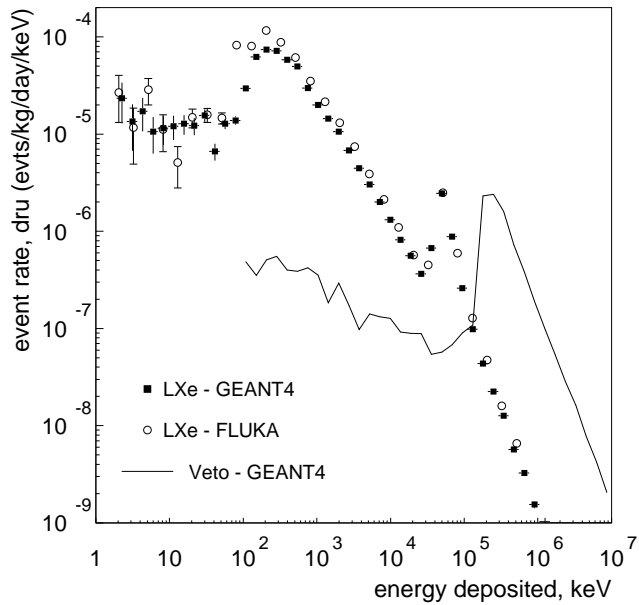}
   \caption{Differential spectra of the total energy deposited in the
   liquid xenon (LXe) target as predicted by GEANT4 and FLUKA and in
   the veto scintillator according to GEANT4 (the latter is scaled
   down by a factor of $5\times 10^4$).}
   \label{edep-em}
\end{figure}

The total energy deposition spectra in the xenon target according to
both MC codes are plotted in Fig.~\ref{edep-em}. This is mainly
`electromagnetic' energy: nuclear recoils have a noticeable
contribution only at the lowest energies. The simulations agree within
statistical errors in the low-energy region below 100~keV. The
discrepancy observed near 100~keV may be due to isotope-production
followed by radioactive decay with gamma emission, which was modelled
in the FLUKA simulation but not considered in this GEANT4 work. FLUKA
predicts more gammas than GEANT4 between 100~keV and a few tens of MeV
--- which is not surprising considering that it also produces more
neutrons --- but the general agreement is still reasonable. At higher
energies, where the passage of primary muons causes the peak seen in
the figure, the two spectra nearly coincide. The spectrum of the total
energy deposited in the veto is also shown (uncorrected by the
quenching factor in the case of proton recoils).

\begin{figure}
   \includegraphics[width=8.5cm]{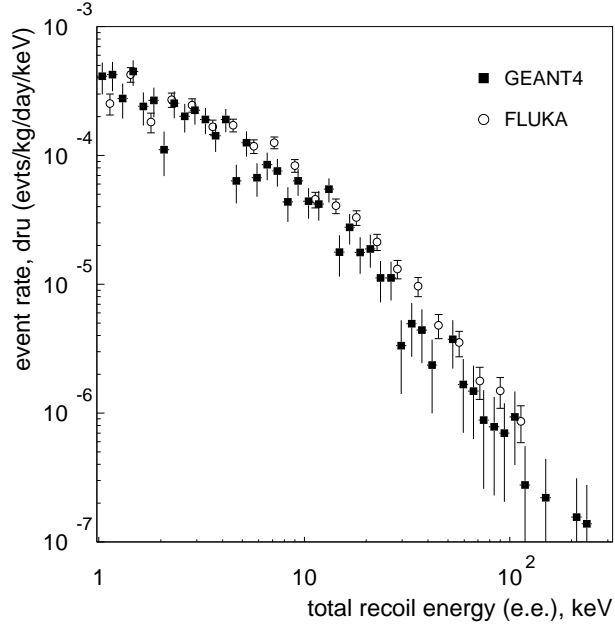}
   \caption{Nuclear-recoil energy spectrum in the liquid xenon
   detector as a function of the visible energy deposited by all
   nuclear recoils in each event. The spectra include `mixed' events
   involving electromagnetic energy deposits, not just `pure' nuclear
   recoils, but only the energy left by nuclear recoils was counted.}
   \label{recoils}
\end{figure}

To obtain a statistically significant dataset of nuclear recoil
events, an exposure time of nearly 3 years was simulated with GEANT4
($\simeq 20\times 10^6$ muons). FLUKA simulations were performed with
the new release FLUKA-2003 giving the statistics of about 4~years of
live time. The nuclear recoil spectrum in the xenon target is shown in
Fig.~\ref{recoils}. The spectra include both `pure' and `mixed' recoil
events, the latter term representing events involving both
electromagnetic energy and nuclear recoils. The total number of
recoils with any energy predicted by GEANT4 for the 250~kg target is
217$\pm$9 per year, of which 10.3$\pm$1.9 are not accompanied by
additional energy deposits. In the visible energy range of 2--10~keV,
where most WIMP signals should occur, 3.3$\pm$1.1 recoils are expected
per year. A similar figure of 3.9$\pm$1.0 is predicted by FLUKA. The
results from the two codes are compared in Table~\ref{table1}.

\begin{figure}
   \includegraphics[width=8.5cm]{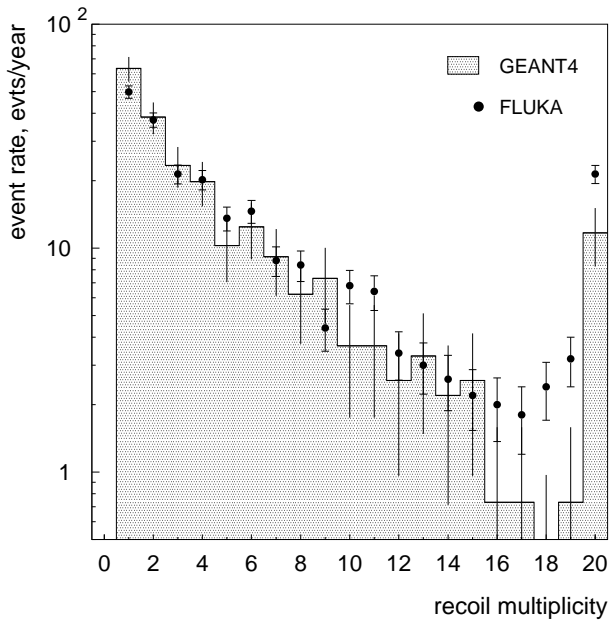}
   \caption{Distribution of the nuclear-recoil multiplicity in the
   250~kg xenon target, for all events containing neutron elastic
   scatters. The last data-point includes multiplicities of 20 and
   above.}
   \label{multirec}
\end{figure}

Note that the numbers presented in Table~\ref{table1} for FLUKA agree
with the previous simulations (Paper~3) for the total recoil rate, but
are a factor of 2 smaller for `pure' nuclear recoil events. This is
due to the lower energy cuts for gammas and electrons used in the
present simulations (50 keV for xenon and its vessel, 100 keV for
hydrocarbon and lead and 1 MeV for rock). This makes the probability
of coincidences between nuclear recoils and gammas/electrons in xenon
higher, and reduces the rate of `pure' nuclear recoils.

A detector with good spatial resolution can resolve the locations of
multiple neutron elastic interactions, which occur on average several
cm apart. These are not regarded as irreducible background since they
should not be caused by WIMPs. The recoil multiplicity in the target
(i.e. the distribution of the number of recoils per MC event) is shown
in Fig.~\ref{multirec} (including all recoil events, of all
energies). GEANT4 predicts that 28\% of these are single recoils.
Considering only those with visible energies greater than 2~keV ---
i.e. which have a realistic chance of being detected --- the fraction
of single scatters decreases to 14\%. Finally, the simulation
registered 2.6$\pm$1.0 single recoils per year in the energy range
2--10~keV~e.e., whereas FLUKA indicates 2.1$\pm$0.7 recoils per
year. In both MCs all recoil events were vetoed with a signal greater
than 100~keV, i.e. no recoils were observed in the target in
anti-coincidence with the veto.

%%%%%%%%%%%%%%%%%%%%%%%%%%%%%%%%%%%%%%%%%%%%%%%%%%%%%%%%%%%%%%%%%%%%%%%%
\begin{table}
\caption{\label{table1}Muon-induced neutron background (nuclear recoil
(NR) events per year) in 250~kg xenon target for several detection
thresholds}
\begin{tabular}{l|c|cc}
\hline
\hline
Event type              & $E_{ee}$, keV & FLUKA        & GEANT4       \\
\hline
All NR events           & $>$0          & 234$\pm$8    & 217$\pm$9    \\
Pure NR events          & $>$0          & 13.6$\pm$1.9 & 10.3$\pm$1.9 \\
                        & $>$2          & 7.3$\pm$1.4  & 4.0$\pm$1.2  \\
                        & 2--10         & 3.9$\pm$1.0  & 3.3$\pm$1.1  \\
Pure single NR events   & $>$0          & 6.6$\pm$1.3  & 6.6$\pm$1.6  \\
                        & 2--10         & 2.1$\pm$0.7  & 2.6$\pm$1.0  \\
Anti-coincidence with veto & $>$0       & 0            & 0            \\
\hline
\end{tabular}
\end{table}
%%%%%%%%%%%%%%%%%%%%%%%%%%%%%%%%%%%%%%%%%%%%%%%%%%%%%%%%%%%%%%%%%%%%%%%%

In spite of predicting different neutron yields, the results
summarised in Table~\ref{table1} suggest that FLUKA and GEANT4 arrive
at a similar total number of neutron events, as well as similar rates
in the range 2--10~keV~e.e. This is explained by the neutron spectra
at the rock face shown in Fig.~\ref{shielding}: although the {\em
total} yields from FLUKA can be a factor of 2 higher in materials such
as salt and lead, the actual integral flux above 1~MeV, capable of
producing detectable nuclear recoils, differs by about 20--30\%.

\begin{figure}
   \includegraphics[width=8.5cm]{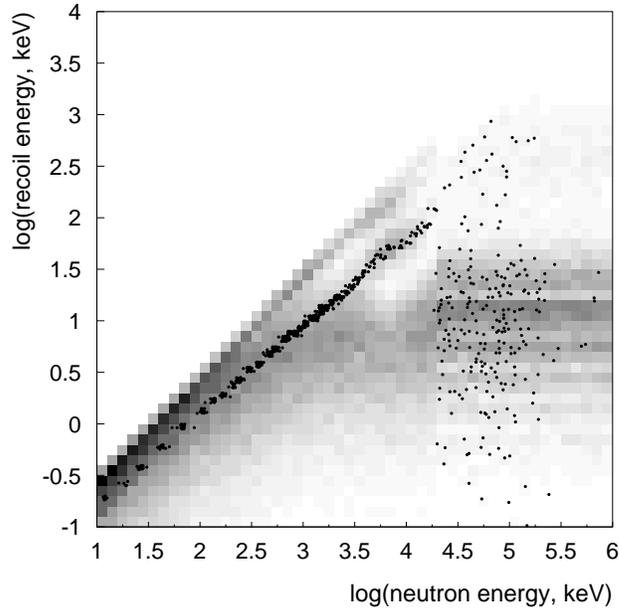}
   \caption{Kinetic energy of xenon recoils as a function of neutron
   energy in the FLUKA simulation (filled markers) and for a
   log-uniform neutron distribution with GEANT4 (grey-scale). In both
   MC codes, neutron elastic scattering is treated by different models
   below and above $\simeq$20~MeV.}
   \label{nelastic}
\end{figure}

Although the nuclear recoil spectrum is generally harder in FLUKA than
in GEANT4, the number of events at 2--10~keV does not differ much in
the two models. The difference between these spectra is mainly
explained by the different treatment of neutron elastic scattering in
the two codes. The data-based models describing elastic interactions
below 20~MeV are more accurate in GEANT4 than in FLUKA. This is
illustrated in Fig.~\ref{nelastic}, which plots the nuclear recoil
energies in the xenon target obtained with FLUKA against the
corresponding neutron energy. A new dataset was generated with GEANT4
for a log-uniform neutron energy distribution, also shown in the
figure. Both MCs exhibit a discontinuity at $\simeq$20~MeV, where
parameterised models take over from the low-energy ones. The most
striking difference is that FLUKA generates mono-energetic recoils
(based on Kerma factors) for a fixed neutron energy in the low energy
region, whereas GEANT4 is much more accurate, reproducing the full
recoil spectrum.

%%%%%%%%%%%%%%%%%%%%%%%%%%%%%%%%%%%%%%%%%%%%%%%%%%%%%%%%%%%%%%%%%%%%%%%
\section{Conclusions}

We have investigated how GEANT4 and FLUKA model muon-induced neutrons
in bulk matter in an attempt to validate these Monte Carlo codes for
the simulation of the neutron backgrounds expected in underground
laboratories. The total yields, neutron energy spectrum and lateral
distribution from the muon track in hydrocarbon material obtained with
both codes are consistent with the available measurements, with GEANT4
predicting fewer neutrons than FLUKA above $\sim$100~GeV muon energy
by a small factor ($<$30\%). In most other materials studied, GEANT4
also generates lower neutron fluxes for muon energies in the hundreds
of GeV. At 280~GeV, the typical average muon energy at several
underground laboratories, the total neutron yield can be a factor of 2
lower for some materials. These include salt and lead, which are
important as rock and shielding media. In general, GEANT4 shows
enhanced neutron photoproduction in electromagnetic cascades in light
materials, although this tendency is less marked relative to the
recent FLUKA release. For heavy elements, FLUKA produces more neutrons
than GEANT4 in both electromagnetic and hadronic cascades initiated by
muons. A new theory-driven model for the hadronic vertex of the
muon-nucleus interaction will appear in a future GEANT4 release
\cite{hp}, which may increase the hadronic yields. Experimental yields
obtained for Pb \cite{gorshkov74,bergamasco73} appear to be mutually
inconsistent, but may suggest that both simulations underestimate
neutron production in heavy targets. Direct production in muon
spallation reactions is similar in both codes, although the simulated
rates for fast neutrons ($>$10~MeV) disagree with experimental data
obtained in the CERN NA55 experiment. Such marked disagreement is not
observed with measurements by the E665 Collaboration for neutrons
under 10~MeV.

The two packages were used to model the muon-induced background in an
idealised WIMP detector containing 250~kg of liquid xenon as an active
medium. The simulated total fluxes of fast neutrons ($>$1~MeV)
entering the underground cavern were found to agree within
20\%. Similar agreement was obtained for the neutron flux emerging
after 30~cm of lead and 50~cm of hydrocarbon shielding. The nuclear
recoil spectrum observed in the xenon target is slightly harder in the
FLUKA simulation, which also predicts a small excess of `pure' recoil
events. However, both simulations indicate a similar number of recoils
with visible energy of 2--10~keV. A single-recoil rate of the order of
2~events/year is predicted by both models, which is probably not far
below the value expected from an `ideal' Monte Carlo code. We do not
expect a large enhancement in the neutron flux after all possible
improvements to the codes have been implemented.

%%%%%%%%%%%%%%%%%%%%%%%%%%%%%%%%%%%%%%%%%%%%%%%%%%%%%%%%%%%%%%%%%%%%%%%
\section{Acknowledgments}

This work has been undertaken within the framework of the UK Dark
Matter Collaboration (University of Edinburgh, Imperial College
London, Rutherford Appleton Laboratory and University of Sheffield)
and contributes to the collaboration-wide simulation efforts. The work
is supported by the UK Particle Physics and Astronomy Research Council
(PPARC). The authors wish to thanks H. P. Wellisch (CERN) for his
invaluable assistance and for his comments to this manuscript. Thanks
are due to M. Robinson (University of Sheffield) and L. Mendes
(Imperial College) for their help with computational facilities. We
also acknowledge the funding from the EU FP6 project ILIAS.

%%%%%%%%%%%%%%%%%%%%%%%%%%%%%%%%%%%%%%%%%%%%%%%%%%%%%%%%%%%%%%%%%%%%%%%

\end{document}